# TEMPERATURE DEPENDENCE OF GRAVITATIONAL FORCE: EXPERIMENTS, ASTROPHYSICS, PERSPECTIVES

A.L. Dmitriev

St-Petersburg University of Information Technologies, Mechanics and Optics
49, Kronverksky Prospect, St-Petersburg, 197101, Russia. E-mail: dalexl@rol.ru

**Abstract**

*The consistency of the results of measuring the gravitational force temperature dependence obtained by Shaw and Davy in 1923 and by the author in 2003 was shown. Such dependence is observed in the laboratory experiments, it does not contradict the known facts of classical mechanics and agrees with astrophysics data. It was pointed out that experimental research into temperature influence on gravitation was needed and perspectives of developing that trend in gravitation physics was promising.*

The problem of influence of temperature of bodies on their gravitational interaction was naturally raised at the very early period of development of gravitation physics. The first attempts to experimentally determine the relation between temperature and gravitation did not produce any results due to low accuracy of measurements [1]. Late in the XIX-th century, following the development of electromagnetic theories of gravitation predicting an increase in body gravitation force with an increase in their temperature the interest to relevant experiments rose. By 1916 the most accurate measurements of dependence of gravitational force $F(t)$ in the temperature interval of $t$ 20 – 220° C (for big mass),

$$F(t) = F_0(1 + \alpha t), \qquad (1)$$

were carried out by Shaw who obtained positive value of temperature coefficient $\alpha = +(1.20 \pm 0.05) \cdot 10^{-5}$ degree$^{-1}$ [ 2 ]. In 1923 Shaw and Davy pointed out the fallibility of that result and on the basis of measurements carried out with higher accuracy they concluded "that the effect, if it exists, is less than $2 \cdot 10^{-6}$ per degree, and may well be zero" [ 3 ]. Actually, Shaw and Davy obtained a negative value $\alpha = -2.0 \cdot 10^{-6}$ degree$^{-1}$ which is illustrated in Fig. 1.

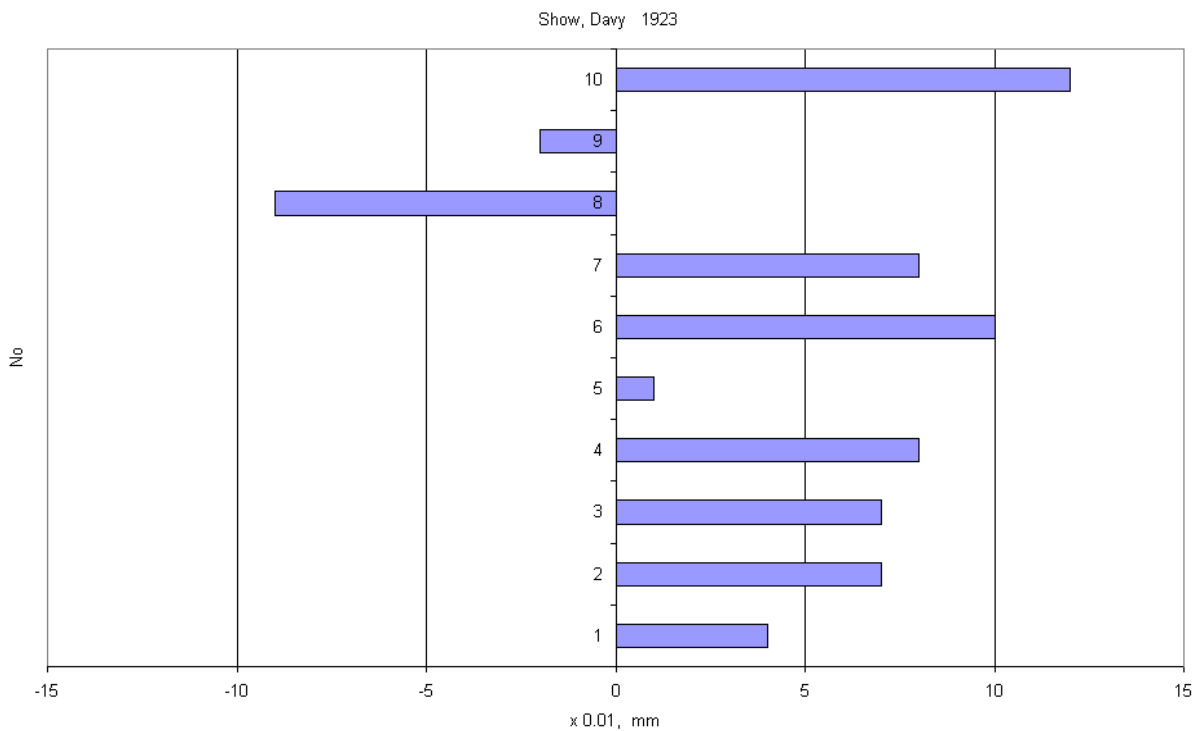

Fig. 1. The generalized results of experiments by Shaw and Davy ([3], Table II). The abscissa corresponds to the difference in the forces of gravity created by heated and cold massive bodies.

Nevertheless, the authors did not insist on the accuracy of the nonzero result which they obtained and regarded it as approximately equal to the measurement error magnitude; the actual measurement error of the average magnitude $\alpha$, according to the data of [3], is less than 40%. Such a conservative estimate probably was caused by the fact that in the early 20ties the general theory of relativity (GTR) began to become rather popular, according to which the temperature effect on gravitation in experiment was practically not observed [4]. The impact of GTR was so great that over next 80 years any experimental research into temperature dependence of the force of gravity was not carried out due to its "nonscientific nature". Incidentally, the basic in GTR principle of equivalence was experimentally checked only with strict stabilization of temperatures of interacting bodies; the direct experimental evidences justifying that principle under different temperatures of test bodies are not available till the present time.

In 2003 our work [5] was published which experimentally confirmed physical dependence of body weight on its temperature. The physical prerequisite of temperature dependence of the force of gravity, according to the phenomenological model, is the dependence of acceleration of the force of gravity on the magnitude and sign of accelerations caused by influence of external elastic forces on the test body [6]. With increase of body temperature the accelerated movement of its component particles becomes more intensive which causes exponential, $\propto T^{1/2}$, dependence of the force-of-gravity acceleration on absolute temperature $T$ of the body [5,7]. An example of experimental dependence of apparent mass of brass rod on duration of its heating (and temperature) is shown in Fig. 2.

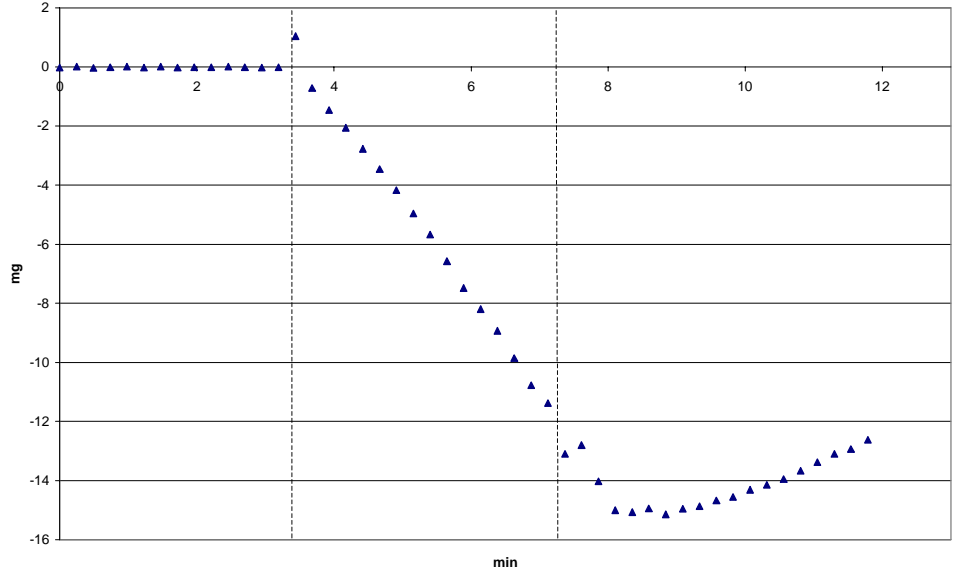

Fig. 2. Change of apparent mass of brass rod in the process of its heating [5].

Temperature dependence of attractive force of two bodies, in the first approximation, can be described by the known Newton's gravity law with gravity constant in the form of

$$G = G_0(1 - a_1\sqrt{T_1})(1 - a_2\sqrt{T_2}),  \quad (2)$$

where $G_0$ – constant, $T_1$ и $T_2$ – absolute temperatures of interacting masses (exceeding Debye temperature), and $a_2$ - temperature coefficients the magnitude of which depends on density and elastic properties of body materials. With a constant temperature of one of the masses (for example, $T_1 =$ const) the coefficients $\alpha$ (1) and $a_2 = a$ (2) are directly proportional,

$$\alpha = -\frac{a}{2\sqrt{T}}, \quad (3)$$

where $T = T_2$ – average temperature of the other mass. According to the experimental estimates the magnitude of coefficient $a$ is minimal for viscous and dense bodies and, for example, for lead and brass, it is close to $a \approx 1.5 \cdot 10^{-4} K^{1/2}$ [5]. For example, with $T = 470K$ for lead we find $\alpha \approx -3.5 \cdot 10^{-6}$ degree$^{-1}$. Obviously, the signs of temperature change of gravitational force in experiments [3] and [5] coincide, and the magnitudes of corresponding temperature coefficients, with account for measurement errors, are close to each other. That's why both Shaw and Davy's of 1923 and our experiments of 2003 give evidence of the laboratory-observed physical temperature dependence of the force of gravity: decrease of attractive force of bodies with increase in their temperature.

It is interesting to consider some astrophysical consequences of the above dependence.

Slow change of the planet average temperature due to either the radiant heat exchange on its surface or to internal heat processes within its volume changes, according to (2), the effective magnitude of gravity constant. As a result, the **planet orbit precesses** with angular deflection of perihelion for one planet revolution equal to

$$\delta\varphi \approx \frac{\pi k P}{2(1-e^2)}, \quad (4)$$

where coefficient $k$ describes changes in time of the average planet temperature,

$$k = -\frac{a}{2\sqrt{T}}\frac{dT}{dt}, \quad (5)$$

$P$ – period of planet turnover, $e$ – orbit eccentricity (taken as $e \ll 1$), $t$ – time [7]. For example, in slow cooling of the planet the force of its gravitational interaction with the sun increases (we take constant the average temperature of the sun), coefficient $k$ is positive and orbit precession is direct.

The next example is a **double pulsar**. Slow cooling of the stars involved in the system of double pulsar determines the increase in the force of their gravitational interaction. As a result, the period of turnover of twin stars is reduced and their orbit periastrons are deflected. Relative change in period $P$ of the double pulsar is equal to

$$\frac{\Delta P}{P} = \frac{1}{4}\left(\frac{a_1 \Delta T_1}{\sqrt{T_1}} + \frac{a_2 \Delta T_2}{\sqrt{T_2}}\right), \quad (6)$$

where $T_1$ and $T_2$ – average temperatures of stars, $a_1$ и $a_2$ – their effective temperature coefficients [7].

While analyzing the complicated **movements of near-to-the sun plasma** it was noticed that in the sun vicinity the effective magnitude of the gravity constant is less than its standard value. Taking into account the fact that the temperature of plasma near the sun is high, in the order of $10^4 - 10^5 K$, the fact is directly explained by the dependence (2).

The measuring of **gravitational constant**. Divergencies of the experimental values of the gravitational constant might be caused by inequality of the absolute temperatures of sample masses used in various gravitational experiments [8].

If the temperature dependence of the force of gravity keeps in extreme processes of **black hole** formation then a similar singularity appears to produce some doubts. Formation of black holes is hindered by the pressure forces on the part of collapsing substance which might exceed those of gravitational compression. In so doing, the compression process slows down and might transform itself into the phase of scattering (heat explosion) of the substance; on the whole, that process might be both of monotone and oscillating in time nature.

So, today there are experimental grounds to consider real the marked dependence of gravitational force on the absolute temperatures of interacting mass. Such dependence does not contradict the known experimental facts of classical mechanics and naturally agrees with the data of astrophysics [7]. Comprehensive research into the temperature dependence of the gravitational forces in the wide range of temperatures (including low ones, as well) of the test bodies of various physical composition will allow in the perspective to establish new peculiarities of the gravitational interaction of those bodies. Physics of gravitation might receive new development similar to that which optics received in transition from heat to laser light sources.